\documentclass{ifacconf}

\usepackage{graphicx}      % include this line if your document contains figures
\usepackage{natbib}        % required for bibliography

\usepackage{amssymb}
\usepackage{graphicx}
\usepackage{url}
\usepackage{xcolor}
\usepackage{amsmath}
\usepackage{mathtools}
\usepackage{tikz}
\usepackage{verbatim}
\usepackage{subcaption}
\usepackage{pgfplots}

\usepackage{graphicx}
\usepackage{epstopdf}
\usepackage{amssymb}
\usepackage{relsize}
\usepackage[textwidth=3.8em,textsize=scriptsize,disable]{todonotes}
\usepackage{algorithm}
\definecolor{Gray}{gray}{0.90}
\usepackage{colortbl}

\usepackage{algorithm}
\usepackage{algorithmicx}
\usepackage{algcompatible}
\usepackage{algpseudocode}

\usepackage{color}
\usepackage{kpfonts}
\usepackage{verbatim}
\usepackage{lipsum}% http://ctan.org/pkg/lipsum

\newtheorem{theorem}{\bf{Theorem}}[section]

\newenvironment{definition}[1][Definition]{\begin{trivlist}
\item[\hskip \labelsep {\bfseries #1}]}{\end{trivlist}}
\def\qed{\hfill\rule[-1pt]{5pt}{5pt}\par\medskip}
\usepackage{amssymb}
\usepackage{url}

\newcommand{\calV}[0]{\mathcal{V}}
\newcommand{\calE}[0]{\mathcal{E}}

\newcommand{\calS}[0]{\mathcal{S}}
\newcommand{\calN}[0]{\mathcal{N}}

\newcommand{\calR}[0]{\mathcal{R}}

\newcommand{\calC}[0]{\mathcal{C}}

\definecolor{blue}{rgb}{0,0,0}
\begin{document}
\begin{frontmatter}

\title{Leveraging Diversity for Achieving Resilient Consensus in
Sparse Networks}%\thanksref{footnoteinfo}} 

%\thanks[footnoteinfo]{Sponsor and financial support acknowledgment
%goes here. Paper titles should be written in uppercase and lowercase
%letters, not all uppercase.}

\author[First]{Faiq Ghawash} 
%\author[Second]{Aritra Mitra} 
\author[First]{Waseem Abbas}

\address[First]{Information Technology University, Lahore, Pakistan \\(e-mails: faiq.ghawash@itu.edu.pk, w.abbas@itu.edu.pk)}
%\address[Second]{Purdue University, West Lafayette, IN USA \\(e-mail: mitra14@purdue.edu)}
%\address[Third]{Electrical Engineering Department, 
%   Seoul National University, Seoul, Korea, (e-mail: author@snu.ac.kr)}

\begin{abstract}                % Abstract of not more than 250 words.
%----- New Abstract -----------
A networked system can be made resilient against adversaries and attacks if the underlying network graph is structurally robust. For instance, to achieve distributed consensus in the presence of adversaries, the underlying network graph needs to satisfy certain robustness conditions. A typical approach to making networks structurally robust is to strategically add extra links between nodes, which might be prohibitively expensive. In this paper, we propose an alternative way of improving network's robustness, that is by considering heterogeneity of nodes. Nodes in a network can be of different types and can have multiple variants. As a result, different nodes can have disjoint sets of vulnerabilities, which means that an attacker can only compromise a particular type of nodes by exploiting a particular vulnerability. We show that, by such a diversification of nodes, attacker's ability to change the underlying network structure is significantly reduced. Consequently, even a sparse network with heterogeneous nodes can exhibit the properties of a structurally robust network. Using these ideas, we propose a distributed control policy that utilizes heterogeneity in the network to achieve resilient consensus in adversarial environment. We extend the notion of $(r,s)$-robustness to incorporate the diversity of nodes and provide necessary and sufficient conditions to guarantee resilient distributed consensus in heterogeneous networks. Finally we study the properties and construction of robust graphs with heterogeneous nodes.    
%----- Old Abstract -----------
%In a networked system, the design of distributed control algorithms for achieving consensus in adversarial environment is a challenging task. Networked systems are made resilient against faults and adversaries by making the underlying network structurally robust. Typically, structurally robustness is improved by strategically adding extra links in the network. However, limited sensing range of the participating agents allows communication among close proximity neighbours making connectivity augmentation impossible or prohibitively expensive. In this paper, we aim to improve the robustness of the networked system by introducing heterogeneous set of nodes in the network. Heterogeneity means the nodes in the network are of different types and have disjoint set of vulnerabilities. We show that, heterogeneity in the network significantly reduces attackers ability to change the underlying network allowing a sparse network to exhibit the properties of structurally robust network. We propose a distributed control policy that utilizes the heterogeneity in the network to achieve resilient consensus in adversarial environment. We extend the notion of $(r,s)$-robustness to incorporate the diversity of nodes in the network and provides necessary and sufficient conditions in heterogeneous networks to achieve resilient consensus in adversarial environment. Finally we study the properties and present methods for the construction of robust heterogeneous networks.      
\end{abstract}


\end{frontmatter}
%===============================================================================
\section{Introduction}
A key aspect of networked and cooperative systems is that participating agents rely on local interactions to achieve complex global tasks such as area coverage, network formation and flocking (for instance, see \cite{mesbahi2010graph,ren2005survey}). The success of various distributed control algorithms depend on agents sharing true information with each other and updating their states according to the predefined update protocols. However, if a small subset of agents do not adhere to the designed control protocols and share incorrect information with neighbors, the overall performance of the system is adversely impacted, and the overall network objective might not be achieved. For instance, in the distributed consensus problem, which is one of the most widely studied problem in distributed computing and network control systems, a single malicious node can prevent all the other nodes from converging to a common state, which is the overall network objective. Consequently, resilience of distributed control algorithms to malicious agents is an important issue.

A typical approach to improving resilience of distributed algorithms in networks is to exploit the underlying network structure. In fact, it has been shown that highly connected and structurally robust networks are more resilient to adversarial intrusions. For instance, to achieve distributed consensus, there are algorithms that guarantee consensus in the presence of misbehaving nodes if the underlying network graph satisfies certain connectivity and robustness conditions, for instance  \cite{leblanc2013resilient,park2017fault,tseng2015fault}. A common aspect in all these solutions is that the underlying network must satisfy certain robustness and connectivity conditions, which ultimately result in highly dense graphs. Although these solutions perform well, the high connectivity requirements might limit the scope of these solutions, especially in sparse networks. For instance, in many practical scenarios, where the communication links are formed based on proximity, it might be difficult or prohibitively expensive to establish new communication links between agents. Thus, an interesting question is how can we improve the resilience of such distributed algorithms in sparse networks? In other words, how can we make sparse networks \emph{act like} highly connected or robust networks without adding communications links?

In this paper, we propose an alternative way of improving network's robustness, that is by utilizing the notion of \emph{diversity} of nodes. In simple words, diversity means that nodes in a network are of different types and have many variants. Nodes might be different based on their hardware implementation, software, resources etc. A consequence of having diverse implementations is that nodes might have disjoint vulnerability sets, and by exploiting a particular vulnerability, an attacker can only impact the nodes belonging to that particular type. Thus, diversification effectively limits the attacker's ability to compromise a large number of nodes within the network, and hence improves the network's robustness. We modify a typically used notion of network robustness (as in \cite{leblanc2013resilient}) to include the effect of nodes' diversity, and then utilize it to design a resilient distributed consensus algorithm that guarantees convergence even in sparse networks. Our main contributions are:

\begin{itemize}
\item We formalize the notion of \emph{$(r,s)$-robustness with coloring} that takes into account the effect of nodes' diversity by assigning colors to nodes. Then, we show that the robustness properties of the graph, even a sparse one, can be significantly improved by appropriately introducing diversity of nodes instead of adding extra edges with in the network.

\item We propose a resilient distributed consensus algorithm and determine conditions---in terms of $(r,s)$-robustness with coloring---that guarantee consensus among normal nodes in the presence of adversaries.

\item We also discuss the construction of $(r,s)$-robust graphs with colors. Finally, we provide simulations that verify our results.
\end{itemize}

%------------------- Para ---------
The concept of diversification of nodes has been employed previously in computer networks as an effective security mechanism (e.g. see \cite{o2004achieving,newell2015increasing,alarifi2006diversify}). Apart from diversification of nodes, an alternative approach to improve the structural robustness is by making few carefully chosen nodes as trusted (immune to attacks by hardening), for instance \cite{dziubinski2013network,abbas2019td}. Trusted nodes can help to design distributed algorithms that achieve resilient consensus (\cite{abbas2018improving}) and resilient estimation (\cite{mitra2018impact}), however, the assumption of guaranteeing the true operation of trusted nodes at all times is highly optimistic, and may require large investment to harden such agents.%n hardware and software hardening of such agents to achieve their trusted status.
%----------------------------------

The rest of the paper is organized as follows: Section \ref{sec:prob} provides a network model and formulates the problem. Section \ref{diversity} discusses the diversity paradigm and introduces the notion of $(r,s)$-robustness with coloring. Section \ref{sec:algo} presents a resilient distributed consensus algorithm with a detailed analysis. Section \ref{sec:construct} discusses construction of robust heterogeneous networks. Section \ref{sec:simulation} provides simulation results, and finally, Section \ref{sec:con} concludes the paper. 

%============================================================================== 

\section{Network Model and Problem Formulation} \label{Problem Formulation}
\label{sec:prob}
%\subsubsection{Multi-Agent System Modelling}
A multi-agent system is modelled by an undirected graph $\mathcal{G(V,E)}$. The vertex set $\mathcal{V}$ corresponds to agents (e.g robots, sensors), whereas the the edge set $\mathcal{E}$ represents the communication model among the agents. An edge between node $i$ and $j$ shows the information exchange between nodes and is represented by $(i,j)=(j,i)$. The \emph{neighbourhood} of node $i$ is defined as $N(i)= \{ j \in \mathcal{V} : (i,j)\in \mathcal{V}\} $, and the \emph{closed neighborhood} is $N[i] = N(i) \cup \{i\}$. We use the terms nodes, vertices and agents interchangeably. At any time instant $t$, each node has a state value denoted by $x_{i}(t)\in\mathbb{R}$. Based on the application, the state value can be a sensor measurement, position variable, optimization parameter, opinion or any other quantity of interest. We assume that all nodes interact synchronously with each other. Our network is heterogeneous in the sense that nodes are of different types. 

\subsubsection{Node types} There are multiple types (or variants) of nodes in the network, and each node belongs to a specific type. We denote node types by a set of colors $\Gamma = \{C_1,C_2,\cdots,C_n\}$. All nodes of type $i$ are assigned color $C_i$. Nodes can be of different types based on their hardware platforms, software, resources, or due to other implementation or functional features. Moreover, each node in the network is either \emph{normal} or \emph{adversarial} as defined below:
\subsubsection{Normal nodes}
A normal node is the one that always updates its state according to a predefined update rule based on the values of nodes in $N[i]$, for instance,
\begin{equation}
\label{eq:basic}
x_{i}(t+1) = f(\{x_{j}(t)\}), \: j\in N[i].
\end{equation}
 
\subsubsection{Adversarial nodes}
Adversarial nodes are the ones that are compromised by an attacker (for instance, by exploiting vulnerabilities), and therefore, do not follow the state update rule  \eqref{eq:basic}. They can change their state values arbitrarily.  %Non-cooperative behaviour may occur due component malfunctioning, privacy concerns or the node may gets compromised by the adversary. 
We consider that all adversarial nodes must belong to the \emph{same type}, that is, they all have the \emph{same color}. Adversarial nodes may feed others with malicious and misleading information, thus preventing the network to achieve the required global objective. We note that a normal node knows the colors of its neighbors, but does not know the type (color) of adversarial nodes.

\subsubsection{Threat models and scopes}
If an adversary node shares the same state value at time $t$ with all of its neighbors, then it is commonly known as a \textit{malicious adversary}. Similarly, an adversarial node sending different values to different nodes in its neighbors at time $t$ is commonly referred to as the \textit{byzantine adversary}.
The scope of the threat is usually defined in terms of the number of adversarial nodes in the network, for instance using \emph{F-total} and \emph{F-local} models \cite{leblanc2013resilient}. In our case, $F$-total model means that there are at most $F$ adversarial nodes (of the same color) in the network. Similarly, $F$-local model means that the neighborhood of any node in the network contains at most $F$ adversarial nodes of the same color.

\subsection{Objective: Resilient Consensus}
Here, our objective is to design a distributed control policy for the normal agents such that \textit{agreement} and \textit{safety} conditions are satisfied in the presence of adversarial agents. Agreement condition requires the asymptotic convergence of normal nodes state values to a common value (consensus), whereas safety condition requires that at all times, the state value of any normal node is within the interval defined by the maximum and minimum of the initial values of normal nodes. More precisely, to achieve resilient consensus, the following conditions must be satisfied: 
\begin{enumerate}
\item As $t\to \infty$, $x_{i}(t) = x_{j}(t) = x $ for all normal nodes $i$,$j$.
\item Let $M[t]$ and $m[t]$ denotes the maximum and minimum values of normal nodes at any time step $t$. For all normal nodes $i \in \mathcal{V}$, $m[0] \leq x_{i}(t) \leq M[0]$ 
\end{enumerate}

\section{Diversity for Improving Network's Structural Robustness}
\label{diversity}
Heterogeneity in networks has been studied in many different contexts by researchers across various disciplines. One such aspect is the diversification of nodes, which broadly means that nodes in a network are of different types, and have multiple variants. Our goal is to exploit diversity of nodes to effectively improve the network's structural robustness. In other words, we explore if it is possible to limit an attacker's ability to change the underlying network structure by having a variety of nodes. Diversification of nodes can be achieved by employing different operating systems, software packages, and hardware platforms. Owing to distinct implementations of such variants, they typically have disjoint exploitation sets and vulnerabilities. As a result, an attacker cannot compromise devices of different types (with disjoint vulnerabilities) by exploiting a particular vulnerability at a time. In fact, an attacker can only compromise devices belonging to the same type or class by exploiting a particular vulnerability specific to that class. This effectively limits attackers ability to attack nodes in the network. Thus, if we assign colors to nodes depending on the particular type they belong to, then the attacker can only attack nodes of the same color by exploiting the vulnerabilities corresponding to that particular type. Building on this simple yet key observation, we model the diversity of nodes to improve network's robustness in adversarial environment. 

%As discussed in the earlier section,the nodes in the network are of different types and have multiple variants. 
We consider that each node belongs to a specific node type, which is represented by a color, and each node is then assigned a unique color. 

\begin{definition}[Definition 1](\textit{Coloring}) Let $\Gamma = \{C_{1},C_{2}, \dots C_{n}\}$ be the set of colors, then \emph{coloring $\mathcal{C}$} %$\mathcal{C}$ 
is the assignment of colors from $\Gamma$ to nodes in $\mathcal{V}$, that is 
\begin{equation}
\label{eq:coloring}
\mathcal{C}: \; \cal{V} \; \longrightarrow \; \Gamma
\end{equation}
The number of colors used in the heterogeneous network is represented by $C$. For the ease of notation, we denote the color assigned to node $i$ by $\calC_i$.
\end{definition}

\begin{definition}[Definition 2](\textit{Mono-chromatic and Poly-chromatic Subsets}) A subset of nodes $S\subseteq \cal{V}$ where $|S|>1$ is mono chromatic, if all nodes in $S$ have the same color, that is $\calC_{i} = \calC_{j}$, $\forall i,j\in S$. Otherwise $S$ is poly-chromatic. 
\end{definition}

%\begin{definition}[Definition 3](\textit{Poly-chromatic nodes}) A subset of nodes $S\subseteq V$ is poly-chromatic, if nodes in $S$ have different colors. 
%\end{definition}

% $(r,s)$-robustness has been an important metric in determining the resilience of distributed algorithm (e.g WMSR). The $(r,s)$-robustness of network is defined as follows.

% \begin{definition}[Definition 4] ($(r,s)$-robustness) Let $S_1$ and $S_2$ be arbitrary non-empty, disjoint subsets. Let $\mathcal{X}_{S_1}^{r}$ and $\mathcal{X}_{S_2}^{r}$ be the set of nodes that has r-neighbours in the set $\mathcal{V} \setminus S_{i}$ where i $\in \{1,2\}$ respectively. A graph $\mathcal{G(V,E)}$ is $(r,s)$-robust if at least one of the following condition is satisfied:

% \begin{itemize}
% \item [(i)] %all nodes in $S_1$ are $r$-valid nodes, (that is 
% $|\mathcal{X}_{S_1}^{r}| = |\mathcal{S}_1|$
% \item [(ii)] %all nodes in $S_2$ are $r$-valid nodes, (that is 
% $|\mathcal{X}_{S_2}^{r}| = |\mathcal{S}_2|$
% \item [(iii)] $|\mathcal{X}_{S_1}^{r} \cup \mathcal{X}_{S_2}^{r}|\ge s$
% \end{itemize}

% \end{definition}

To measure network robustness and quantify the effect of adversarial nodes on the overall performance of the network, we utilize the notion of \emph{$(r,s)-robustness$} as in \cite{leblanc2013resilient}. This notion has turned out to be very useful in analyzing the resilience of distributed algorithms, in particular distributed consensus, in adversarial set-ups. Next, we modify the notion of $(r,s)$-robustness to incorporate the diversification of nodes. 
\begin{definition}[Definition 3](\textit{r-valid node}) For a positive integer $r$ and a subset $S\subset \cal{V}$, a node $v\in S$ is an \emph{r-valid node} if at least one of the following is satisfied:
\begin{itemize}
\item[(i)] $v$ has at least $r$ mono-chromatic neighbors outside of $S$ (that is, in $N(v)\setminus S$).
\item[(ii)] $v$ has at least two neighbors with different colors outside of $S$.
\end{itemize}
\end{definition}
%An illustration of above conditions is shown in Figure \ref{Fig1}.

% \begin{figure}[h]
% \begin{subfigure}[b]{0.3\textwidth}
% \centering
% \includegraphics[scale=0.75]{validity_1_new}
% \caption{}
% \end{subfigure} \quad \quad
% \begin{subfigure}[b]{0.1\textwidth}
% \centering
% \includegraphics[scale=0.75]{validity_2}
% \caption{}
% \end{subfigure}
% \caption{\small{Illustration of conditions (i) and (ii) for a node $v$ to be $r$-valid.}}
% \label{Fig1}
% \end{figure}

%\begin{definition}[Definition 4] (\textit{$r$-robustness with coloring})
%A graph is \emph{$r$-robust with coloring}, if for any pair of non-empty disjoint subset $S_{1}$,$S_{2}$ $\subset V$, at least one of the subset must contain one $r$-valid node.
%\end{definition}

\begin{definition}[Definition 4]
\label{def:rs}
(\textit{$(r,s)$-robustness with coloring}) Let $r,s>1$ be two positive integers, and $S_1$ and $S_2$ be non-empty, disjoint subsets of $\calV$. Let $\mathcal{X}_{S_1}^{r}$ and $\mathcal{X}_{S_2}^{r}$ be the set of $r$-valid nodes in $\mathcal{S}_1$ and $\mathcal{S}_2$ respectively. A graph $\mathcal{G(V,E)}$ is \emph{ $(r,s)$-robust with coloring} if at least one of the following is always satisfied:
\end{definition}

\begin{itemize}
\item [(i)] %all nodes in $S_1$ are $r$-valid nodes, (that is 
$|\mathcal{X}_{S_1}^{r}| = |\mathcal{S}_1|$.
\item [(ii)] %all nodes in $S_2$ are $r$-valid nodes, (that is 
$|\mathcal{X}_{S_2}^{r}| = |\mathcal{S}_2|$.
\item [(iii)]  $(\mathcal{X}_{S_1}^{r} \cup \mathcal{X}_{S_2}^{r})$ is mono-chromatic and $|\mathcal{X}_{S_1}^{r} \cup \mathcal{X}_{S_2}^{r}|\ge s$.
\item [(iv)]  $(\mathcal{X}_{S_1}^{r} \cup \mathcal{X}_{S_2}^{r})$ is poly-chromatic. 
\end{itemize}

\begin{figure}[h]

\begin{subfigure}[b]{0.24\textwidth}
\centering
\includegraphics[scale=0.15]{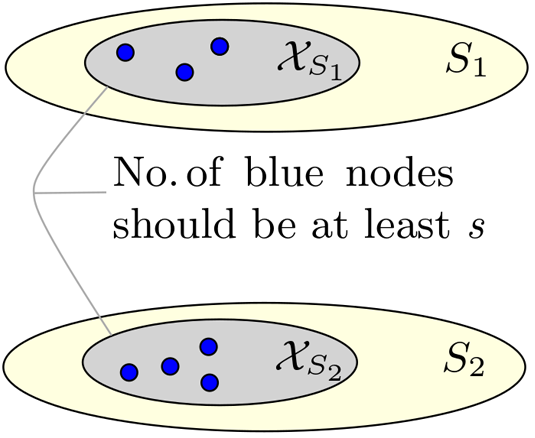}
\caption{}
\end{subfigure} \quad \quad
\begin{subfigure}[b]{0.2\textwidth}
\centering
\includegraphics[scale=0.15]{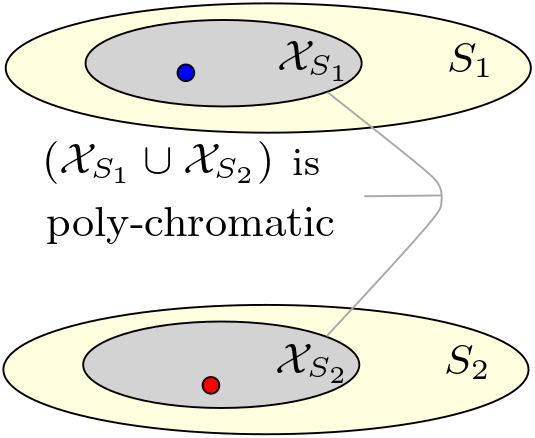}
\caption{}
\end{subfigure}
\caption{\small{Conditions (iii) and (iv) for $(r,s)$-robustness with coloring.}}
\label{Fig2}
\end{figure}

The condition (iv) above requires at least two distinct colored $r$-valid nodes in $S_{1} \cup S_{2}$. We note that if all nodes in the network are of the same color, then the above definition is exactly the notion of $(r,s)$-robustness defined in \cite{leblanc2013resilient}. An illustrations of conditions (iii) and (iv) are given in Figure \ref{Fig2}.

% \begin{figure}[h]
% \includegraphics[scale=1.1]{Conditions_Comb.png}
% \caption{Illustration of conditions (iii) and (iv) of the $(r,s)$-robustness with coloring.}
% \label{fig:validity}
% \end{figure}
%%%----------------Figure of Examples Section -----------------%%%%%%

The main idea here is that a graph that is $(r,s)$-robust (with only one color) can be $(r',s')$-robust with (multiple) colors for $r'>r$ and $s'>s$. In other words, it is possible to make sparse networks highly robust with colors assigned appropriately to nodes. We also note that $(r,s)$-robustness can only be improved using colors if the underlying graph is at least $(2,2)$-robust. Similarly, we define the notion of \emph{$r$-robustness with coloring}, which will be used to analyze the resilient consensus in the case of $F$-local adversary model, as below: 

\begin{definition}[Definition 5] (\textit{$r$-robustness with coloring})
A graph is \emph{$r$-robust with coloring}, if for any pair of non-empty disjoint subset $S_{1}$,$S_{2}$ $\subset V$, at least one of the subsets must contain a node that has at least $r$-neighbours of any color or at least three distinct color neighbours outside of its respective set.
\end{definition}

It must be noted that $r$-robustness can only be improved using colors if the underlying graph is at least $3$-robust. We also note that the notion of $r$-robustness with coloring and $(r,s)$-robustness with coloring differ with respect to the validity of node having poly-chromatic neighbourhood. Hence $(r,1)$-robustness with coloring does not correspond to $r$-robustness with coloring except in networks with only one type of nodes. We explain the significance of colors and the notion of $(r,s)$-robustness with coloring in the examples below.
%Note that, it directly follows from the above definitions that, $r$-robustness of the graph can only be improved, by introducing heterogeneity in the graph when the underlying graph topology is at least $2$-robust. Similarly, a graph must have to be at least $(2,2)$-robust to utilize the heterogeneity in the network.

\subsection{Examples}
\begin{figure}[hb]
\centering
\begin{subfigure}[b]{0.15\textwidth}
\centering
\includegraphics[scale=0.23]{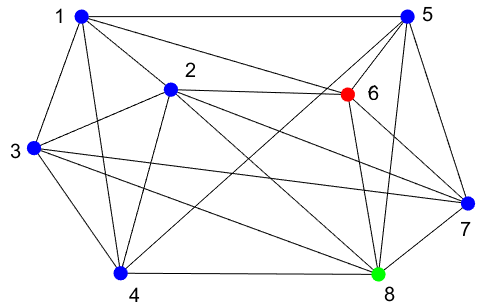}
\caption{}
\end{subfigure}
\begin{subfigure}[b]{0.15\textwidth}
\centering
\includegraphics[scale=0.23]{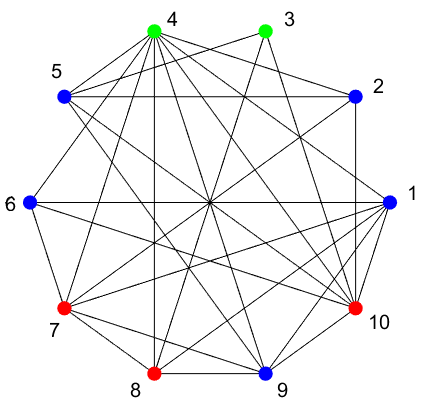}
\caption{}
\end{subfigure}
\begin{subfigure}[b]{0.15\textwidth}
\centering
\includegraphics[scale=0.23]{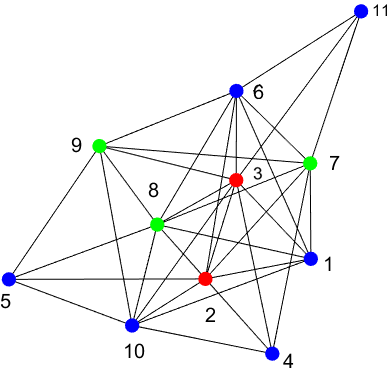}
\caption{}
\end{subfigure}
\caption{\small{(a) $(4,4)$-robust graph with colors. (b) $(5,5)$-robust graph with colors. (c) $5$-robust graph with colors.}}
\label{Fig3}
\end{figure}

The graph in Figure \ref{Fig3}(a) is $(2,2)$-robust with one color. However, if we use three colors such that nodes in the set $\{1,2,3,4,5,7\}$ are assigned color $C_1$, node 6 is assigned $C_2$, and node 8 has color $C_3$, then the graph becomes $(4,4)$-robust with three colors. Similarly, the graph in \ref{Fig3}(b) is not $(3,3)$-robust if all the nodes have the same color. By using three colors, the graph becomes $(5,5)$-robust, for instance, if nodes in $\{1,2,5,6,9\}$ have color $C_1$, nodes in $\{7,8,10\}$ have $C_2$ and nodes in $\{3,4\}$ are assigned $C_3$. Finally, the graph in Figure \ref{Fig3}(c) is $3$-robust which becomes $5$-robust with three colors, where nodes in the sets $\{1,4,5,6,10,11\}, \{2,3\}$ and $\{7,8,9\}$ have colors $C_1,C_2$ and $C_3$ respectively.

%%%%%%%%%------------Figure----------------#######

\begin{figure}[ht]
\centering
\begin{subfigure}[b]{0.16\textwidth}
\centering
\includegraphics[scale=0.16]{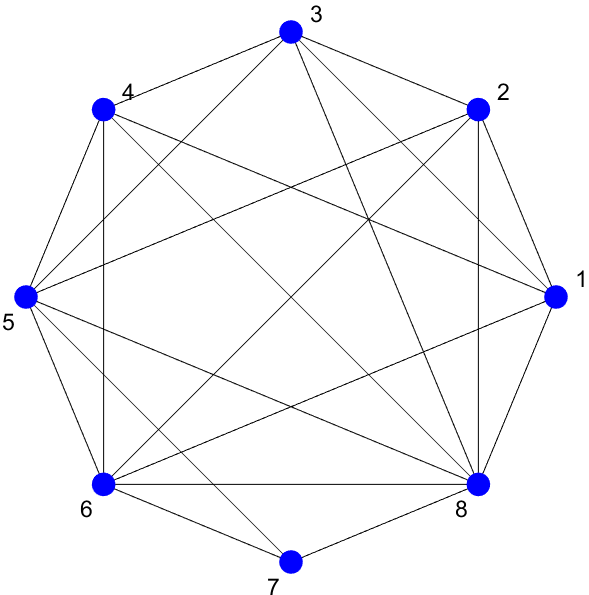}
\caption{}
\end{subfigure}
\begin{subfigure}[b]{0.16\textwidth}
\centering
\includegraphics[scale=0.19]{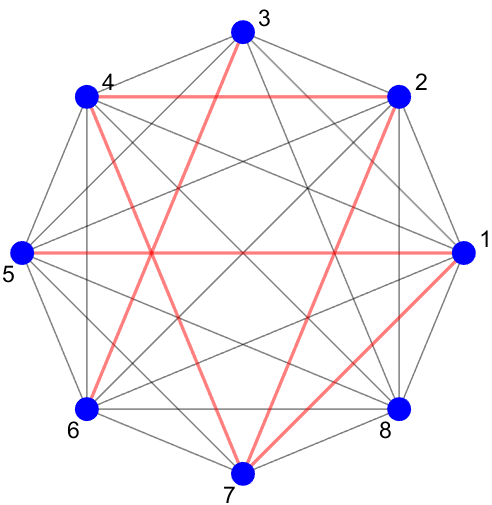}
\caption{}
\end{subfigure}
\begin{subfigure}[b]{0.15\textwidth}
\centering
\includegraphics[scale=0.16]{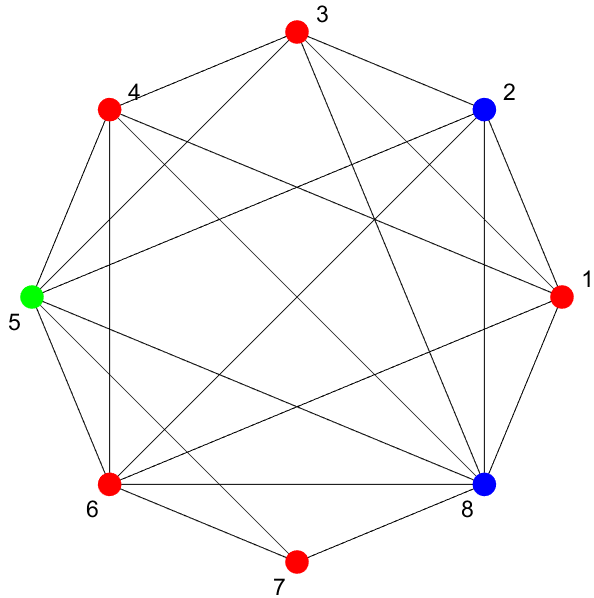}
\caption{}
\end{subfigure}
\caption{\small{(a) A $(2,2)$-robust graph with one color nodes. (b) Six edges must be added to make the graph $(4,4)$-robust. (c) Graph becomes $(4,4)$-robust by appropriately introducing three colors.}}

\label{Fig4}
\end{figure}

These examples emphasize that diversity of nodes in the network can be used to significantly improve the robustness of the network. A network can exhibit the properties of highly connected and structurally robust networks without adding additional communication links. Figure \ref{Fig4} shows the comparison of adding additional links to the nodes coloring (diversification) approach. It is observed that diversification could improve robustness in networks, especially in cases where adding communication links are prohibitively expensive or is not feasible. Moreover, it is interesting to note that multiple coloring schemes can be utilized to achieve the desired robustness.% This allows us to utilize the available resources effectively to achieve the desired robustness in the network. 

%%%%%%%%%------------Figure----------------#######

\section{Resilient Consensus Protocol with Coloring (RCP-C)}
\label{sec:algo}
In this section, we present a distributed state update rule for normal nodes, which we term as \textit{Resilient Consensus Protocol with Coloring} (RCP-C) that guarantees consensus among normal nodes in heterogeneous networks under certain robustness conditions discussed later. The main idea is that, at any time step $t$, node $i$ collects values from all of its neighbors, but considers only a subset of them to update its state. We note that every node knows the colors of its neighbors, but does not know the color of adversarial nodes. The values of neighbors considered by $i$ are explained below:

%First, a node always considers values that are obtained after removing $F$ largest (and $F$ smallest) values with respect to node's own value $x_i(t)$. Here, $F$ is a given parameter, and is an upper bound on the number of adversaries (under the $F$-total or $F$-local models). Second, based on the colors of neighbors, node $i$ groups values that are greater (smaller) than $x_i(t)$. Then, $i$ ignores all the values in the group containing the maximum (minimum) value, and considers all the values in the remaining groups. 

First, a node determines $F$ largest and $F$ smallest values corresponding to nodes in $N_i(t)$. Let such nodes be denoted by $\overline{\calR}_i(t)$ and $\underline{\calR}_i(t)$ respectively. The node $i$ always considers values of nodes in $N_i(t)\setminus(\overline{\calR}_i(t)\cup\underline{\calR}_i(t))$. Here, $F$ is a given parameter, and is an upper bound on the number of adversaries (under the $F$-total or $F$-local models). Second, based on the colors of nodes in $\overline{\calR}_i(t)$ (respectively $\underline{\calR}_i(t)$), node $i$ groups values into various subsets. Then, $i$ ignores all the values in the subset containing the maximum (minimum) value, and considers all the values in the remaining subsets.

\begin{figure}
\centering
\includegraphics[scale=0.9]{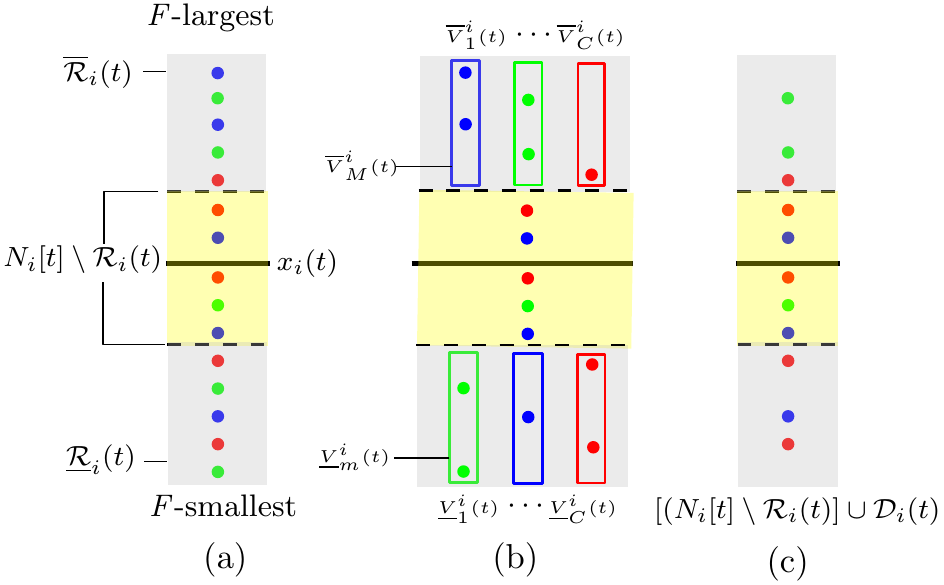}
\caption{\small{Illustration of steps followed in RCP-C. (a) Node $i$ computes $F$ largest (and $F$ smallest) values with respect to node's own value (Steps 1 and 2). (b) Based on the colors of neighbors, node $i$ groups values in $\overline{\calR}_i(t)$ (respectively $\underline{\calR}_i(t)$) and ignores all the values in the group containing the maximum (minimum) value (Step 3). (c) All remaining values in (a) and (b) are considered in the update rule of node $i$ (Step 4).}}
\label{fig:algo}
\end{figure}

Next, we formally present the steps in the algorithm below, and illustrate in Figure \ref{fig:algo}.
  
\begin{enumerate}
\item At each time step $t$, node $i$ receives state values from its neighbours ${N}_{i}(t)$. 

\item The neighbouring nodes ${N}_{i}(t)$ are categorized into two sets $\overline{N}_{i}(t)$ and  $\underline{N}_{i}(t)$ based on their state values as follows:\\
\begin{center}
$\overline{N}_{i}(t)$  =$ \{ l \in {N}_{i}(t)\: : \:  x_{l}(t) >  x_{i}(t) \}$ \\
$\underline{N}_{i}(t)$ = $ \{ l \in {N}_{i}(t)\: : \:  x_{l}(t) <  x_{i}(t) \}$\\
\end{center}
Next, define $\overline{\mathcal{R}}_{i}(t)$ = $\overline{N}_{i}(t)$ if $|\overline{N}_{i}(t)|<F$. Otherwise, $\overline{\mathcal{R}}_{i}(t)$ consists of the $F$ nodes in $\overline{N}_{i}(t)$ with the highest state values. Similarly, define $\underline{\mathcal{R}}_{i}(t)$ = $\underline{N}_{i}(t)$  if $|\underline{N}_{i}(t)|<F$. Otherwise $\underline{\mathcal{R}}_{i}(t)$ consists of the $F$ nodes in $\underline{N}_{i}(t)$  with the lowest state values. Finally, we define $$\mathcal{R}_{i}(t) = \underline{\mathcal{R}}_{i}(t) \cup \overline{\mathcal{R}}_{i}(t) $$

\item  Based on the number of colors in the neighborhood of node $i$, $\overline{\mathcal{R}}_{i}(t)$ is divided into different subsets, $\{\overline{V}^i_1(t),\overline{V}^i_2(t),\cdots \overline{V}^i_C(t)\}$, where $\overline{V}^i_k(t)$ contains the values in $\overline{\mathcal{R}}_{i}(t)$ corresponding to nodes with color $C_k$. Consider $\overline{V}^i_M(t)$ to be the subset containing the maximum value in $\overline{\mathcal{R}}_{i}(t)$.\footnote{Ties are broken arbitrarily.} Moreover, define 
$$\overline{\mathcal{D}_{i}}(t) =\overline{\mathcal{R}}_{i}(t) \setminus \overline{V}^i_M(t).
$$ \\
Similarly, divide $\underline{\mathcal{R}}_{i}(t)$ into subsets, $\{\underline{V}^i_1(t), \cdots \underline{V}^i_C(t)\}$, and consider $\underline{V}^i_m(t)$ to be the subset containing the minimum value in $\underline{\mathcal{R}}_{i}(t)$. Then, define 
$$\underline{\mathcal{D}_{i}}(t) =\underline{\mathcal{R}}_{i}(t) \setminus \underline{V}^i_m(t).
$$ \\
Finally, we define $$\mathcal{D}_{i}(t) = \overline{\mathcal{D}}_{i}(t) \cup \underline{\mathcal{D}}_{i}(t).$$

\item Each normal node \textit{i} updates its value according to the following rule:
\begin{equation}
\label{eq:res_consensus}
x_{i}(t+1) = \sum_{j\in[({N}_{i}[t] \setminus \mathcal{R}_{i}(t)]\cup \mathcal{D}_{i}(t)} w_{ij}(t)x_{j}(t) 
\end{equation}
\end{enumerate}

Here, $w_{ij}(t)$ represents the weight that is assigned to the value of node $j$ by node $i$ at time step $t$ \footnote{$\sum_{j=1}^{n}w_{ij}=1$, where $w_{ij} \geq \alpha$ for some $0< \alpha < 1.$}. An illustration of various steps of the algorithm is given in Figure \ref{fig:algo}. 

The RCP-C differs from the conventional Weighted Mean Subsequent Reduced (WMSR) algorithm (\cite{leblanc2013resilient}) in step-3. This step allows the algorithm to utilize the diversity of nodes in the network and extract useful information based on the colors (types) of nodes in the neighbourhood. 

\subsection{Analysis}
Next, we analyze RCP-C algorithm, and provide necessary and sufficient conditions to guarantee resilient consensus in the presence of adversaries ($F$-total and $F$-local models) in heterogeneous networks. The main results are stated below. 

\begin{theorem}
\label{analysis_theorem}
Let $\mathcal{G(V,E)}$ be a time invariant heterogeneous network, in which each node is assigned a color from the coloring set $\Gamma = \{C_1,C_2,\cdots,C_n\}$, and each normal node follows RCP-C. Then,

\begin{enumerate}
    \item under the \textit{F-total malicious model}, resilient asymptotic consensus is achieved if and only if the underlying graph topology is $(F+1,F+1)$-robust with colors. 
    \item Similarly, under the \textit{F-local malicious model}, 
    resilient asymptotic consensus is achieved if the underlying graph topology is $(2F+1)$-robust with colors.
    %$(F+1)$-robustness with colors is a necessary condition, whereas
    %$(2F+1)$-robustness with colors is a sufficient condition to achieve resilient asymptotic consensus.
    
\end{enumerate}

The proof of Theorem \ref{analysis_theorem} is given in Appendix \ref{appen:A}.

\end{theorem}
%================== New Section ====================
\section{Construction and Properties of Heterogeneous Robust Graphs}
\label{sec:construct}
In this section, first we discuss the construction of $(r,s)$-robust graphs with colors. Since the exact computation of $(r,s)$-robustness is computationally challenging, even if all the nodes are of the same color (as discussed in \cite{zhang2015notion}), it is useful to develop approaches to grow networks by adding nodes while preserving robustness.

\begin{theorem}
\label{growing_graph}
Let $\mathcal{G(V,E)}$ be an $(r,s)$-robust graph with colors. Then the graph $\mathcal{G}'(\calV \cup \{u_{new}\},\calE')$ obtained by adding a new vertex $u_{new}$ to $\mathcal{G(V,E)}$ is also $(r,s)$-robust with colors if any of the following holds.
\begin{enumerate}
    \item $u_{new}$ is adjacent to at least $r+s-1$ mono-chromatic nodes. 
    \item $u_{new}$ is adjacent to at least $\max(r,s)$ nodes of color $C_{k}$ and one node of any other color $C_{j}$, $j\ne k$.
    \item $u_{new}$ is adjacent to at least three distinct color nodes.
\end{enumerate}
\end{theorem}
The proof of Theorem \ref{growing_graph} is given in Appendix \ref{appen:B}.

Similarly, it can be shown that the property of $r$-robustness with colors remain preserved if a new node is adjacent to $r$ nodes of any color, or it is adjacent to three distinct color nodes in the existing graph.

Next, we analyze the robustness conditions that guarantee consensus of normal nodes implementing RCP-C if \emph{all} the nodes of the same color are compromised. Here, note the difference with the earlier attack model in which at most $F$ nodes of the same color could be adversaries (under the $F$-total or $F$-local set-up). For this, we need to introduce the notion of \emph{mono-chromatic robust graphs}. 

\begin{definition}[Definition 6] (\textit{Mono-chromatic robust})
A graph $\mathcal{G(V,E)}$ with coloring $\mathcal{C}$ is mono-chromatic robust, if for any pair of non-empty disjoint subset $S_{1}$,$S_{2}$ $\subset \calV$, at least one of the subsets contains a node that has at least three distinct color neighbours outside of its respective set.
\end{definition}

Since every normal node has neighbors with multiple colors in a mono-chromatic robust graph, it always considers a value from a normal neighbor to update its state. A node will always have normal neighbor as all the adversarial nodes are of the same color. This gives us the following:

\begin{theorem}
\label{mono-chromatic_robust}
Under RCP-C all normal nodes will reach resilient asymptotic consensus in the presence of any number of malicious adversaries of the same color if the underlying graph topology is mono-chromatic robust.
\end{theorem}
Proof of Theorem \ref{mono-chromatic_robust} is given in Appendix \ref{appen:B}.

Further, we analyze the requirement on minimum number of colors necessary to achieve mono-chromatic robust graphs and provide sharp bounds for some specific graph classes that can be made mono-chromatic robust by a careful assignment of colors to nodes. These graph classes include $F$-elemental graphs (discussed in \cite{guerrero2017formations}) that are inherently $2F+1$-robust and $3$-robust graphs with certain conditions on the neighbourhood of each node.  

\begin{theorem}
\label{min_colors}
Given a graph $\mathcal{G(V,E)}$ with coloring $\mathcal{C}$, at least five colors ($C=5$) are required to make the graph mono-chromatic robust. For $F$-elemental graphs ($F>2$) the bound is sharp.
\end{theorem}
Proof of Theorem \ref{min_colors} is given in Appendix \ref{appen:B}.

\begin{theorem}
\label{3_robust}
A $3$-robust graph in which at least three nodes in the neighbourhood of each vertex are pairwise adjacent, then the number of colors required to make such $\mathcal{G(V,E)}$ mono-chromatic robust is upper bounded by the chromatic number\footnote{Minimum number of colors assigned to nodes such that no two adjacent nodes have the same color.} of $\mathcal{G(V,E)}$.
\end{theorem}
Proof of Theorem \ref{3_robust} is given in Appendix \ref{appen:B}.

Based on the above results, a mono-chromatic robust graph of $n$ nodes can be constructed by starting with a complete graph on five nodes $(K_{5})$ graph and assigning each vertex a unique color. New nodes are added in the network by connecting them with three distinct color nodes in the existing network.
%=================== New Section =====================
\section{Simulation Results}
\label{sec:simulation}
\begin{figure}[ht]
\centering
\begin{subfigure}[b]{0.23\textwidth}
\centering
\includegraphics[scale=0.4]{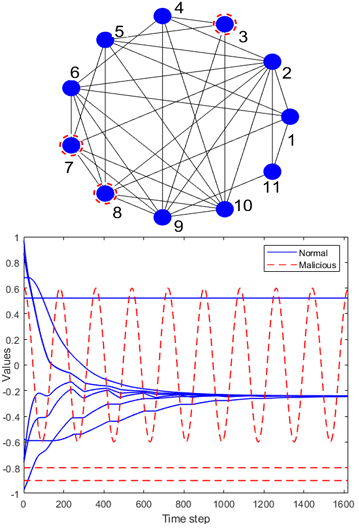}
\caption{$F$-total model (WMSR)}
\label{Fig6a}
\end{subfigure}
\begin{subfigure}[b]{0.2\textwidth}
\centering
\includegraphics[scale=0.4]{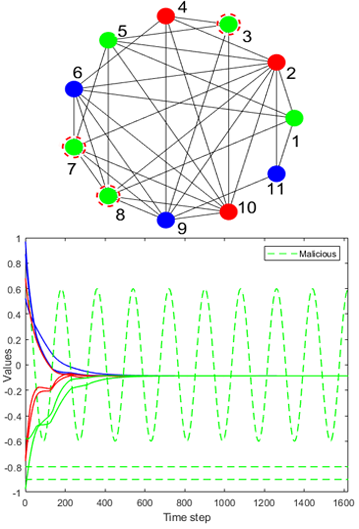}
\caption{$F$-total model (RCP-C)}
\label{Fig6b}
\end{subfigure}
\begin{subfigure}[b]{0.23\textwidth}
\centering
\includegraphics[scale=0.4]{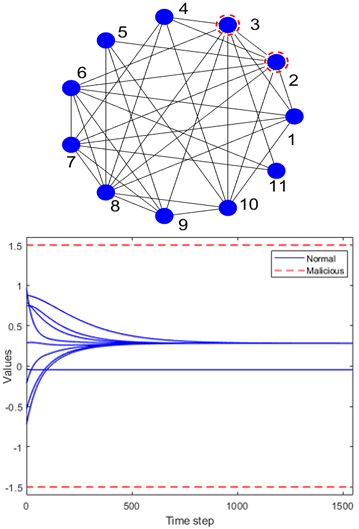}
\caption{$F$-local model (WMSR)}
\label{Fig6c}
\end{subfigure}
\begin{subfigure}[b]{0.2\textwidth}
\centering
\includegraphics[scale=0.4]{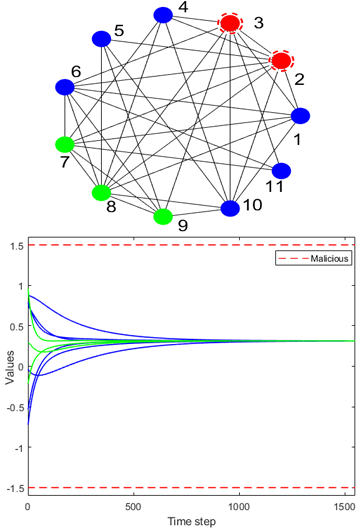}
\caption{$F$-local model (RCP-C)}
\label{Fig6d}
\end{subfigure}
\caption{\small{Comparison of WMSR and RCP-C under the $F$-total and $F$-local malicious models.}} %(a) Under $F$-total model, consensus is not achieved in the homogeneous network implementing WMSR (b) Consensus is achieved under RCP-C.}
\label{Fig6}
\end{figure}

To validate our proposed RCP-C algorithm in heterogeneous networks, we provide simulations for both $F$-total and $F$-local malicious models and compare our approach with existing WMSR algorithm in homogeneous networks. Under the $F$-total malicious model, we consider the network shown in Figure \ref{Fig6}(a) which is $(2,2)$-robust with one color. It means only one malicious node can be tolerated (using WMSR algorithm). Thus, if nodes $3,7,8$ are malicious, consensus is not achieved as shown in Figure \ref{Fig6}(a). However, by an appropriate assignment of colors to nodes (as shown in Figure \ref{Fig6}(b)), the same network becomes $(4,4)$-robust with three colors and can handle up to three malicious nodes. If normal nodes implement RCP-C, consensus is guaranteed even with an attack of three malicious nodes as illustrated in Figure \ref{Fig6}(b). Similarly, the network considered for F-local model simulation is shown in Figure \ref{Fig6}(c). The network is $3$-robust with one color and can tolerate at most 1 malicious node in the neighbourhood of any node. If we consider $F=2$, and nodes 2 and 3 to be malicious, consensus is not achieved under WMSR algorithm. However, the same  network becomes $5$-robust with three colors as shown in Figure \ref{Fig6}(d). Under RCP-C, normal nodes achieve resilient asymptotic consensus as illustrated in Figure \ref{Fig6}(d).  
% Sections and subsections are supported  

% %---------- New Figure ------------------
% \begin{figure*}[h!]
% \begin{subfigure}[t]{0.5\textwidth}
% \centering
% \includegraphics[scale=1.8]{total_WMSR_Finalized.png}
% \caption{homogeneous network implementing WMSR}
% \label{Fig5a}
% \end{subfigure}
% \begin{subfigure}[t]{0.5\textwidth}
% \centering
% \includegraphics[scale=1.8]{total_RCPC_Finalized.png}
% \caption{heterogeneous network implementing RCP-C}
% \label{Fig5b}
% \end{subfigure}
% \caption{Resilient consensus with three malicious nodes (e.g $\{3,7,8\}$) in the network under the \textit{F-total Model} (a) Consensus is not achieved in the homogeneous network implementing WMSR. (b) Consensus is achieved in a the heterogeneous network implementing RCP-C.}
% \label{Fig5}
% \end{figure*}

% %---------- New Figure ------------------
% \begin{figure*}[ht!]
% \centering
% \begin{subfigure}[b]{0.5\textwidth}
% \centering
% \includegraphics[scale=1.2]{Local_WMSR_Finalized_New3.png}
% \caption{homogeneous network implementing WMSR}
% \label{Fig6a}
% \end{subfigure}
% \begin{subfigure}[b]{0.47\textwidth}
% \centering
% \includegraphics[scale=1.2]{Local_RCPC_Finalized_New3.png}
% \caption{heterogeneous network implementing RCP-C}
% \label{Fig6b}
% \end{subfigure}
% \caption{Resilient consensus with two malicious nodes (e.g $\{2,3\}$) in the network under the \textit{F-local Model} (a) Consensus is not achieved in the homogeneous network implementing WMSR. (b) Consensus is achieved in the heterogeneous network implementing RCP-C.}
% \label{Fig6}
% \end{figure*}

%=================== New Section =====================

\section{Conclusions}
\label{sec:con}
This paper proposed an alternative way to improve structural robustness in networks by incorporating the diversification of nodes. We showed that the attacker's ability to change the underlying network could be significantly reduced by deploying diverse nodes. This could effectively lead to a higher robustness in networks, even if they are sparse originally. To account for the robustness of heterogeneous network, we proposed the notion of $(r,s)$-robustness with coloring. We studied the resilient consensus problem and proposed a distributed algorithm that took into account the diversity of nodes in the network and provided conditions in terms of $(r,s)$-robustness with coloring to guarantee consensus in the presence of adversaries. In future, we would like to generalize the attack model by allowing multiple types of nodes to be attacked. Moreover, assigning appropriate types (colors) to nodes to achieve desired robustness is computationally challenging, and we aim to provide efficient algorithms for this problem.
%========== ACKNOWLEDGEMENTS ==============
\section{Acknowledgments}
Authors would like to thank Aritra Mitra and Shreyas Sundaram at Purdue University for helpful discussions.
{\small{
\bibliography{ifacconf}  
}}
%===================== Appendix ===============

%===================== Appendix ===============
\newpage

\appendix
\section{Proofs}    
\label{appen:A}
\subsection{Proof of Theorem \ref{analysis_theorem}  (a)}
In order to prove the theorem, we use similar arguments used in the proof of Theorem 1 in \cite{leblanc2013resilient}.

\emph{\textit{Proof}}: \textit{(\textbf{Necessity}) Under the F-total malicious model $(F+1,F+1)$-robustness with coloring is a necessary condition.}

Let us consider a graph $\mathcal{G}$ that is not $(F+1,F+1)$-robust with coloring. Hence there would exist some non-empty disjoint subsets $S_{1}$ and $S_{2}$ which do not satisfy any of the conditions in Definition 4. Therefore, there would be at most $F$ nodes in $S_{1} \cup S_{2}$ that have two distinct color nodes or $F+1$ neighbours outside of their respective set ($|\mathcal{X}_{S_1}^{F+1} \cup \mathcal{X}_{S_2}^{F+1}|\leq F$). Moreover, we also know that ($\mathcal{X}_{S_1}^{F+1} \cup \mathcal{X}_{S_2}^{F+1}$ ) is mono-chromatic otherwise condition (iv) in Definition 4 would be satisfied. 
As there can be $F$ adversaries in the network under the $F$-total model, we assume that all valid nodes $(\mathcal{X}_{S_1}^{F+1} \cup \mathcal{X}_{S_2}^{F+1})$  are malicious. Moreover, we know that none of the conditions in Definition 4 is satisfied by $\mathcal{G}$. Hence for some $i \in \{1,2\}$ $|\mathcal{X}_{S_i}^{F+1}|<|S_{i}|$, this implies that there would exist at least one normal node in both $S_{1}$ and $S_{2}$, say $x_{1}$ and $x_{2}$, that has at most $F$ mono-chromatic neighbours outside of its respective set. Now consider that, all nodes in $S_{1}$ have state values $a$ and state value in $S_{2}$ be $b$, where $b>a$. The state values of all nodes in $\mathcal{V}\setminus(S_{1} \cup S_{2})$ are assigned values  in the interval $(a,b)$. Malicious nodes keep their values constant throughout. Both $x_{1}$ and $x_{2}$ ignores all values ($F$ or less) outside of their respective set. Hence consensus would not be achieved.

(\textit{\textbf{Sufficiency}) $(F+1,F+1)$-robustness with coloring is a sufficient condition under F-total malicious model.}

Let $\mathcal{A}$ denotes the adversarial nodes in the network, then $\mathcal{N} = \mathcal{V} \setminus \mathcal{A}$ corresponds to the set of normal nodes. We define $M[t] = max_{i\in \mathcal{N}} x_{i}(t)$ and $m[t] = min_{{i\in \mathcal{N}}} x_{i}(t)$. We know that all nodes in $(N_i[t]\setminus \mathcal{R}_{i}(t))\cup\mathcal{D}_{i}(t)$ contains values in the interval $[m[t],M[t]]$ and the update rule is defined as the convex combination of values in the interval. We deduce that $m[t]$ and $M[t]$ are monotone and bounded functions of $t$ and thus both have some limits denoted by $A_{m}$ and $A_{M}$ respectively. In order to achieve consensus among the normal agents we need to show that $A_{M}=A_{m}$.\\
%%%%----------New Line ------%%%%%%%
We will assume that $A_{M}>A_{m}$ and then show that such an assumption leads to contradiction  allowing us to prove $A_{M} = A_{m}$. Let $A_{M}>A_{m}$  and define a constant $\epsilon_{o}$ such that $A_{M}-\epsilon_{o}>A_{m}+\epsilon_{o}$. At any time instant t and for any positive number $\epsilon_{i}$, we define
\begin{equation}
\mathcal{S}_{M}(t,\epsilon_{i}) = \{ j \in \mathcal{V}: x_{j}(t)>A_{M}-\epsilon_{i}\}     
\end{equation}
\begin{equation}
\mathcal{S}_{m}(t,\epsilon_{i}) = \{ j \in \mathcal{V}: x_{j}(t)<A_{m}+\epsilon_{i}\}    
\end{equation}

$\mathcal{S}_{M}(t,\epsilon_{i})$ have all nodes whose state values are greater than $A_{M}-\epsilon_{i}$. Similarly $\mathcal{S}_{m}(t,\epsilon_{i})$ have nodes with values less than $A_{m}+\epsilon_{i}$. It must be noted that $\mathcal{S}_{M}(t,\epsilon_{i})$ and $\mathcal{S}_{m}(t,\epsilon_{i})$ contain both the normal and malicious nodes. Now let $\mathcal{X}_{M}^{F+1}(t,\epsilon_{i}) \subseteq \mathcal{S}_{M}(t,\epsilon_{i})$ be a subset of valid nodes, that is, each node in $\mathcal{X}_{M}^{F+1}(t,\epsilon_{i})$ have $F+1$ mono-chromatic neighbours or two distinct color neighbours outside of $\mathcal{S}_{M}(t,\epsilon_{i})$. Similarly $\mathcal{X}_{m}^{F+1}(t,\epsilon_{i}) \subseteq \mathcal{S}_{m}(t,\epsilon_{i})$ be the subset of valid nodes in $\mathcal{S}_{m}(t,\epsilon_{i})$.\\%A figure can be added\\
Now fix $\epsilon < \frac{\alpha^{N}}{1-\alpha^{N}}$ where $N=|\mathcal{N}|$ denotes total number of normal nodes and here $\epsilon_{o}>\epsilon>0$. 
From the definition of convergence, we know that there exist a $t_{\epsilon}$ such that for any time instant $t>t_{\epsilon}$, $M[t]$ and $m[t]$ are bounded by $A_{M}+\epsilon$ and $A_{m}-\epsilon$.\\
Consider non-empty disjoint subsets $\mathcal{S}_{M}(t_{\epsilon},\epsilon_{o})$ and $\mathcal{S}_{m}(t_{\epsilon},\epsilon_{o})$. From the definition of $\epsilon_{o}$, we note that $\mathcal{S}_{M}(t_{\epsilon},\epsilon_{o})$ and $\mathcal{S}_{m}(t_{\epsilon},\epsilon_{o})$ are disjoint. Since the graph $\mathcal{G}$ is $(F+1,F+1)$-robust with coloring and there can be at most $F$ adversaries in the network. Hence there would always exist one normal valid node in $\mathcal{X}_{M}(t_{\epsilon},\epsilon_{o}) \cup \mathcal{X}_{m}(t_{\epsilon},\epsilon_{o})$. Without loss of generality, we assume that such a valid node (say $i$) is in $\mathcal{X}_{M}(t_{\epsilon},\epsilon_{o})\cap \mathcal{N}$. %Hence minimum value of such a node is  $A_{M}-\epsilon_{o}$.
Next, we show that 
\begin{equation}
\label{peq1}
    x_{i}(t_{\epsilon}+1) \leq A_{M}-\epsilon_{1}, \: \epsilon_{1}< \epsilon_{o}
\end{equation}
In order to compute $x_{i}(t_{\epsilon}+1)$, node $i$ consider values of nodes in the set $(N[k] \setminus \calR_{i}(t))\cup D_{i}(t)$ (as defined in RCP-C Algorithm). %To determine $N[k] \setminus R_{i}(t)$, node $i$ remove $F$ largest (and $F$ smallest) state values with respect to $x_{i}(t)$. Moreover, to compute $D_{i}(t)$, based on the colors of neighbors, node $i$ groups values that are greater (smaller) than $x_{i}(t)$. Then, $i$ ignores all the values in the group containing the maximum (minimum) value,and considers all the values in the remaining groups. 
As the node $i$ has at least $F+1$ mono-chromatic nodes or two distinct color neighbours outside of the set with values less than its own. Thus node $i$ would always consider a value lesser than its own while computing $x_{i}(t_{\epsilon}+1) $. The maximum value of such a neighbour is $A_{M} -\epsilon_{o}$ as it lies in $\mathcal{V}\setminus \mathcal{S}_{M}(t_{\epsilon},\epsilon_{o})$. The maximum value that node $i$, receives from its neighbours in $\mathcal{S}_{M}(t_{\epsilon},\epsilon_{o})$ is $M[t_{\epsilon}]$. Since the update rule is the convex combination of the state values of the neighbours and each combination coefficient is lower bounded by $\alpha$. In the worst case, assigning the maximum weight to the highest value we get 
\begin{center}
 $x_{i}(t_{\epsilon}+1) \leq (1-\alpha) M(t_{\epsilon})+ \alpha(A_{M}-\epsilon_{o})$   
 $$ \leq (1-\alpha) (A_{M}+\epsilon)+ \alpha(A_{M}-\epsilon_{o})$$
 $$ \leq A_{M}-\alpha \epsilon_{o}+(1-\alpha)\epsilon $$
 $$ = A_{M}-\epsilon_{1}$$
\end{center}
where $\epsilon_{1} = \alpha \epsilon_{o} - (1-\alpha)\epsilon$ and $\epsilon_{1}<\epsilon_{o}$. We can repeat the same steps if node $i \in \mathcal{X}_{m}^{F+1}$. Hence
\begin{equation}
\label{peq2}
    x_{i}(t_{\epsilon}+1)\geq A_{m}+\epsilon_{1}
\end{equation}
As a consequence of the \ref{peq1} and \ref{peq2} at least one of the following is true
\begin{itemize}
    \item The number of normal nodes in $\calS_{M}(t+1,\epsilon_{1}) \cap \mathcal{N}$ is strictly lesser than the normal nodes in $\calS_{M}(t,\epsilon_{o}) \cap \mathcal{N}$ e.g $|\calS_{M}(t+1,\epsilon_{1})\cap\mathcal{N}|< |\calS_{M}(t,\epsilon_{o})\cap\mathcal{N}|$.
    \item  $|\calS_{m}(t+1,\epsilon_{1})\cap\mathcal{N}|<|\calS_{m}(t,\epsilon_{o})\cap\mathcal{N}|$.
    
\end{itemize}
Note that $\calS_{M}(t+1,\epsilon_{1})$ and $\calS_{m}(t+1,\epsilon_{1})$ are disjoint as $\epsilon_{1}<\epsilon_{o}$. Next, we define $\epsilon_{j}=\epsilon_{j-1}-(1-\alpha)\epsilon$ for any $j \geq 1$. Note that $\epsilon_{j}<\epsilon_{j-1}$. Then for any time step $t_{\epsilon}+j$, the above analysis can be repeated as long as $\calS_{M}(t_{\epsilon}+j,\epsilon_{j})$ and $\calS_{m}(t_{\epsilon}+j,\epsilon_{j})$ contain normal nodes. Since the number of normal nodes are finite, there exist a time step $t_{\epsilon}+k$ such that at least one of the following is always satisfied:
\begin{enumerate}
    \item[(a)] $\calS_{M}(t_{\epsilon}+k,\epsilon_{k}) = \emptyset$, which implies that the maximum value of any normal node at time step $t_{\epsilon}+k$ is upper bounded by $A_{M}-\epsilon_{k}$, or 
    \item[(b)] $\calS_{m}(t_{\epsilon}+k,\epsilon_{k})=\emptyset$, which implies that the minimum value of the normal nodes is lower bounded by $A_{m}+\epsilon_{k}$.
\end{enumerate}
If  $\epsilon_{k}>0$, then (a) implies a contradiction to the fact that $M[t]$ converges monotonically to $A_{M}$ and (b) contradicts to the fact that the $m[t]$ converges monotonically to $A_{m}$. Next, we show that $\epsilon_{k}>0$.
$$
    \epsilon_{k} = \alpha \epsilon_{k-1} - (1-\alpha)\epsilon = \alpha^{k}\epsilon_{o}-(1-\alpha^{k})\epsilon
    %\geq \alpha^{N}\epsilon_{o}-(1-\alpha^{N})\epsilon
$$
\begin{equation}
    \geq \alpha^{N}\epsilon_{o}-(1-\alpha^{N})\epsilon
\end{equation}
Since $\epsilon< \frac{\alpha^{N}}{1-\alpha^{N}}$, we get $\epsilon_{k}>0$ which gives the desired contradiction, thus proving that $A_{M}=A_{m}$. \qed

\subsection{Proof of Theorem \ref{analysis_theorem} (b)}

\emph{\textit{Proof}}:  (\textit{\textbf{Sufficiency}) $2F+1$-robustness with coloring is a sufficient condition under the $F$-local malicious model.}\\
We can construct the sufficiency proof using the same approach and arguments as followed in proof of the Theorem \ref{analysis_theorem} (a). 
 Recall $\mathcal{N}$ denotes the set of all normal nodes. For $F$-Local malicious model, it must be noted that when considering the non-empty disjoint subsets  $\mathcal{S}_{M}(t_{\epsilon},\epsilon_{o}) \cap \mathcal{N}$ and $\mathcal{S}_{m}(t_{\epsilon},\epsilon_{o}) \cap \mathcal{N}$ defined in the proof of Theorem \ref{analysis_theorem} (a) (the main difference here is that we are considering only normal nodes), at least one of the set contains a node that has at least $2F+1$ or three distinct color neighbours outside of its respective set as the underlying graph is $2F+1$-robust with coloring. Recall that at most $F$ (or less) monochromatic nodes can be compromised, thus there would exist a normal node in ($\mathcal{S}_{M}(t_{\epsilon},\epsilon_{o}) \cap \mathcal{N} \cup \mathcal{S}_{m}(t_{\epsilon},\epsilon_{o})\cap \mathcal{N}) $ that will utilize the value of at least one normal node value outside of $S_{M}(t_{\epsilon},\epsilon_{o})$ or $S_{m}(t_{\epsilon},\epsilon_{o})$ sets. \qed
% By using the same arguments as made in the proof of $F$-total model above, we can show that $2F+1$-robustness with coloring is sufficient condition to achieve consensus in $F$-local model. 

%%%%%%%%%%%%%-------------Appendix B---------%%%%%%%
\section{Proofs}    % Each appendix must have a short title.
\label{appen:B}
\subsection{Proof of Theorem \ref{growing_graph}}
\emph{\textit{Proof}}: Let $S_{1}^{'}$ and $S_{2}^{'}$ be any two non-empty disjoint subsets in $\mathcal{G}^{'}$. For such $S_{i}^{'}$ where $i \in \{1,2\} $ , there can be three cases (a) $u_{new} \not\in S_{i}^{'}$ (b) $\{u_{new}\}= S_{i}^{'}$ (c) $u_{new} \in S_{i}^{'}$.

In the first case, since $\mathcal{G}$ is $(r,s)$-robust with coloring, hence at least one of the condition in Definition 4 is satisfied directly by $S_{1}^{'}$ and $S_{2}^{'}$ in $\mathcal{G}^{'}$.\\
In case (b), under all three clauses of the theorem $u_{new}$ would be a valid node hence condition (i) or (ii) is satisfied in Definition 4.\\
In (c), we can assume that  $u_{new} \in S_{2}^{'}$ without loss of generality. Let $S_{1} = S_{1}^{'} $ and $S_{2} = S_{2}^{'} \setminus \{u_{new}\}$. As the graph $\mathcal{G}$ is $(r,s)$-robust with coloring hence the subset $S_{1}$ and $S_{2}$ satisfy at least one of the conditions in Definition 4. If any of the conditions among (i), (iii) or (iv) is satisfied by $S_{1}$ and $S_{2}$ in graph $\mathcal{G}$ then same condition would be satisfied by $S_{1}^{'}$ and $S_{2}^{'}$ in $\mathcal{G}^{'}$.\\
Now, let us assume condition (ii) is satisfied among all conditions in $\mathcal{G}$ which is $|\mathcal{X}_{S_2}^{r}| = |S_{2}|$. If $|\mathcal{X}_{S_2}^{r}|$ is poly-chromatic then (iv) gets satisfied so we can assume that $S_{2}$ consist of mono-chromatic valid nodes only. Moreover, since only condition (ii) is satisfied among all conditions in Definition 4 hence $|\mathcal{X}_{S_1}^{r}|+|\mathcal{X}_{S_2}^{r}| < s$ and $|\mathcal{X}_{S_2}^{r}| =|S_{2}|$. This implies that $S_{2}$ can have at most $s-1$ nodes. \\
Under the clause (i), if $u_{new}$ in $G^{'}$ is connected to at least $r+s-1$ mono-chromatic nodes, then it must be connected to at least $r$ mono-chromatic nodes outside $S_{2}^{'}$. Similarly under the clause (ii) of the theorem, if $u_{new} \in G^{'}$ is connected to $\max(r,s)$ nodes of $C_{k}$ and one node of $C_{j}$, then $u_{new}$ would be connected to at least $r$ mono-chromatic, or two distinct colors neighbours outside of $S_{2}^{'}$ making $u_{new}$ a valid node. Under the clause (iii) of the theorem, if $u_{new}$ in $\mathcal{G}^{'}$ is connected to three distinct color nodes, it would always be connected to at least two distinct colors nodes outside of $S_{2}^{'}$.\qed

\subsection{Proof of Theorem \ref{mono-chromatic_robust}}
\emph{\textit{Proof}}: The theorem can be proved using the same approach and arguments as followed in proof of the Theorem \ref{analysis_theorem} (a). Note that $\calN$ denotes the set of all normal nodes. For mono-chromatic robust graphs, it must be noted that when considering the non-empty disjoint subsets  $S_{M}(t_{\epsilon},\epsilon_{o}) \cap \mathcal{N}$ and $S_{m}(t_{\epsilon},\epsilon_{o}) \cap \mathcal{N}$ defined in the proof of Theorem \ref{analysis_theorem} (a), at least one of the set contains a node that has at least three distinct color neighbours outside of its respective set. As nodes of only one color are compromised, thus there would exist a normal node in ($\mathcal{S}_{M}(t_{\epsilon},\epsilon_{o}) \cap \mathcal{N} \cup \mathcal{S}_{m}(t_{\epsilon},\epsilon_{o})\cap \mathcal{N}) $ that will utilize the value of at least one normal node value outside of $S_{M}(t_{\epsilon},\epsilon_{o})$ or $S_{m}(t_{\epsilon},\epsilon_{o})$ sets. \qed

\subsection{Proof of Theorem \ref{min_colors}}
\emph{\textit{Proof}}: Let $\mathcal{G(V,E)}$ be a colored $r$-robust graph with four colors ($C=4$).  
Let $S_{1}$ and $S_{2}$ be any arbitrary non-empty disjoint subsets. Without loss of generality, for some $i \in \{1,2\}$ and $j \in \{3,4\}$, we consider $S_{1} = \{l \in \mathcal{V} : \: C(l) \in C_{i}\}$ and $S_{2} = \{l \in \mathcal{V} : \: C(l) \in C_{j}\}$. Then, there does not exist any node in $S_{1}$ or $S_{2}$  which has   three distinct color neighbours outside of its respective set. Hence mono-chromatic robustness can never be achieved with lesser than five colors in the network.

\begin{definition}[Definition 7] (\textit{F-elemental graph})
An $F$-elemental graph is a
graph with $|\mathcal{V}| = 4F +1$ nodes that is $r$-robust with $r = 2F +1$ for some positive integer value of $F$.
\end{definition}

(\emph{Sharpness of bound for $F$-elemental graphs}) Given an $F$-elemental $\mathcal{G(V,E)}$ graph $(F>2)$, the number of vertices are $|\mathcal{V}|=4F+1$ and there exist a set $\mathcal{V}^{'} \subseteq \mathcal{V}$ $(|\mathcal{V}^{'}|=2F)$ such that all nodes in $\mathcal{V}^{'}$ are connected to all vertices in $\mathcal{V}$ \cite[Proposition 1]{guerrero2017formations}. %Moreover, an $F$-elemental graph is $2F+1$-robust. 
Assign five distinct colors to nodes in $\mathcal{V}^{'}$. Then there can be two cases
\begin{itemize}
    \item For some $i \in \{1,2\}$, $S_{i} \cap \mathcal{V}^{'} = \emptyset$: Since each node in $\mathcal{V}^{'}$ is connected to all vertices in $\mathcal{V}$. Then each node in $S_{i}$ is adjacent to five distinct color nodes outside of their respective set allowing all of them to meet the condition in Definition 5.     
    \item For some $i \in \{1,2\}$, $S_{i} \cap \mathcal{V}^{'} \not = \emptyset$: Without loss of generality, we can assume that $S_{1} \cap \mathcal{V}^{'} \not= \emptyset $. If $S_{1}=\mathcal{V}^{'}$ then all nodes in $S_{2}$ have five distinct color neighbours. If $|S_{1} \cap \mathcal{V}^{'}|<|S_{1}|$ then there would exist at least one node in $S_{1}$ or $S_{2}$ that has at least three distinct color nodes outside of its respective sets. 
\end{itemize}
\qed
\subsection{Proof of Theorem \ref{3_robust}}
\emph{\textit{Proof}}: For a graph $\mathcal{G}$ which is $3$-robust (all nodes are of the same color), for every pair of non-empty disjoint subsets $S_{1}$ and $S_{2}$ there exists a node $v$ in $S_{1}\cup S_{2}$ that has at least three neighbours outside of its respective set. Moreover, each vertex in $\mathcal{G}$ has at least three neighbours that are pair wise adjacent. By assigning color to the nodes in the graph such that no two adjacent nodes share the same color (proper coloring), $v$ would always have three distinct color neighbours making $\mathcal{G}$ a mono-chromatic robust graph.\qed

\end{document}